\documentclass[a4paper,11pt]{article}
\usepackage{pos}

\title{Identifying multiwavelength counterparts to astrophysical neutrino events}
 \ShortTitle{Identifying multiwavelength counterparts to astrophysical neutrino events}

\author*[a]{Atreya~Acharyya}
\author[a]{Marcos~Santander}

\affiliation[a]{Department of Physics and Astronomy, University of Alabama, Tuscaloosa, AL 35487, USA}

\emailAdd{aacharyya1@ua.edu}
\emailAdd{jmsantander@ua.edu}

\abstract{High-energy neutrinos originating in astrophysical sources should be accompanied by gamma-rays at production. Depending on the properties of the emission environment and the distance of the source to the Earth, these gamma-rays may be observed directly, or through the detection of lower energy photons that result from interactions with the intervening radiation fields. In this work, we present an automated tool that aims at using data from the \emph{Fermi}-Large Area Telescope to identify multiwavelength counterparts to astrophysical neutrino events. The main goal of this tool is to enable prompt follow-up observations with ground-based and space-based observatories in order to help pinpoint the neutrino source.}

\ConferenceLogo{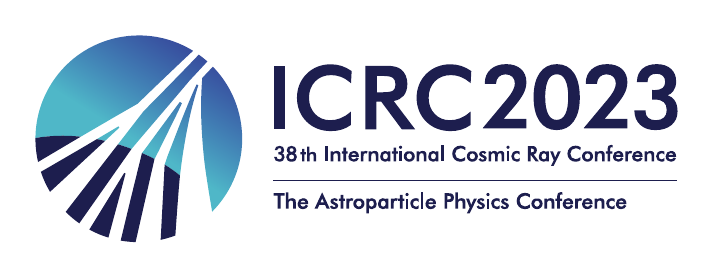}

\FullConference{%
The 38th International Cosmic Ray Conference (ICRC2023)\\
  26 July -- 3 August, 2023\\
  Nagoya, Japan}

\begin{document}
\maketitle

\section{Introduction}

The extragalactic gamma-ray sky is dominated by blazars~\citep{gammaagn}, a subclass of radio-loud active galactic nuclei (AGN) powered by a central supermassive black hole (SMBH), with relativistic jets pointed close to our line-of-sight. The spectral energy distribution (SED) of a typical blazar comprises two distinct peaks. While the first peak, occurring in the radio to the X-ray regime, has been attributed to synchrotron emission from electrons and positrons within the jet, the physical mechanisms responsible for the second peak, produced in the X-ray to gamma-ray regime, is still a matter of debate. Towards understanding this, the \emph{Fermi}-Large Area Telescope (LAT, \cite{Fermi_LAT}) is a pair conversion telescope capable of detecting gamma-ray photons in the energy range between 20~MeV to above 500 GeV. Primarily operating in survey mode, it scans the entire sky every three hours. 

Leptonic models \citep{Blandford_and_Levinson_1995, Georganopoulos_2002} attribute the high-energy peak of the SED to the inverse Compton (IC) scattering of electrons off a source of seed photons, either the same photons emitted through synchrotron emission (synchrotron self-Compton [SSC] model) or photon populations external to the jet (external Inverse Compton [EIC] model). On the other hand, lepto-hadronic models \citep{Romero_2003, MB_2013} suggest that the second peak may be a result of either proton synchrotron or the decay of high-energy mesons produced in cosmic-ray interactions. A hadronic component of the gamma-ray emission~\citep{1993Mannheim} would potentially make AGN prime candidate sources of astrophysical neutrinos \citep{1995Mannheim, 1997Halzen}. 

Since 2016, IceCube has been broadcasting automatic real-time alerts for potential astrophysical neutrino candidate events in order to allow for prompt follow-up observations with ground-based and space-based observatories. The alerts are issued by the Astrophysical Multimessenger Observatory Network (AMON\footnote{\url{https://www.amon.psu.edu} (accessed on 06/06/2023)}, \cite{2013_AMON}) and circulated using the Gamma-ray Coordinates Network (GCN)\footnote{\url{https://gcn.gsfc.nasa.gov} (accessed on 06/06/2023)}, an open-source platform created by NASA to receive and transmit alerts about astronomical transient phenomena. One of the main goals of these alerts is to allow for follow-up observations with other observatories, in the hope of observing multimessenger events. 


In this work, we present an automated tool that aims at using  \emph{Fermi}-LAT data to identify multiwavelength counterparts to astrophysical neutrino events with ground-based and space-based observatories in order to help pinpoint the neutrino source. More specifically, we want to know what is in the region of interest (RoI) around the neutrino alert, whether the RoI is observable from a ground-based observatory and whether anything interesting, for example a gamma-ray flare, is occurring at that particular location. In Section \ref{sec:methods}, we introduce the main components of the analysis and processing pipeline. In Section \ref{sec:discussion}, we consider a typical example of a neutrino alert, IC230506A-gold-rev1, and discuss the primary outputs from the analysis tool. We summarise our conclusions and discuss some plans for future work in Section~\ref{sec:conclusions}.

\section{Methodology}
\label{sec:methods}

\begin{figure*}[t!]
\centering
\includegraphics[width=  0.4 \linewidth]{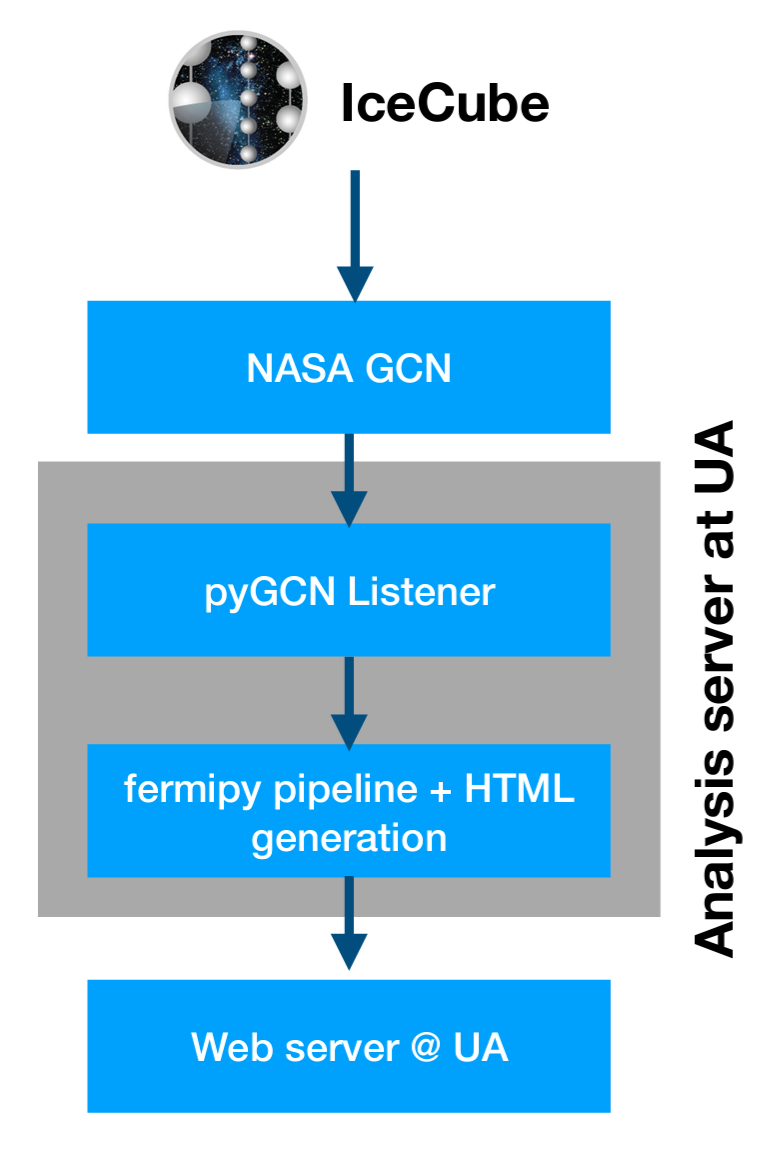}
\caption{A flowchart of the automated pipeline illustrating how each IceCube GCN alert is processed and analyzed before the results are mirrored on to the web server (\url{https://multimessenger.ua.edu/fermi/}, accessed on 06/06/2023).}
\label{fig:2}
\end{figure*}

\begin{figure*}[t!]
\centering
\includegraphics[width=  0.95 \linewidth]{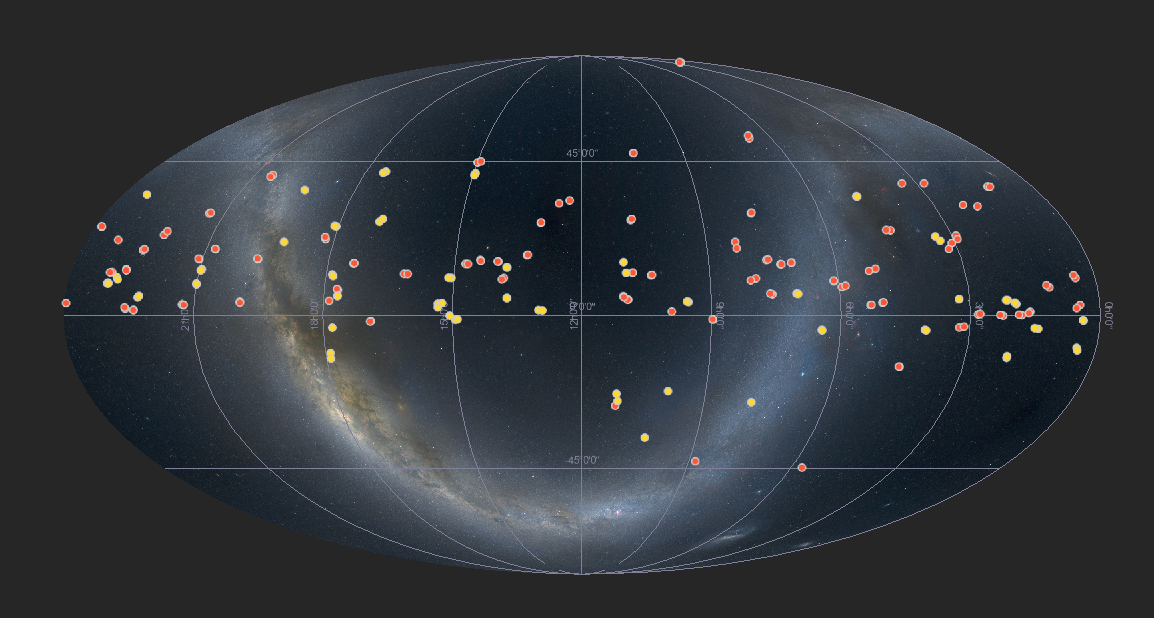}
\caption{A skymap, in celestial coordinates, showing AMON\_ICECUBE\_GOLD events (in yellow) and AMON\_ICECUBE\_ BRONZE events (in red) between June 19, 2019 and June 6, 2023, from the web server (\url{https://multimessenger.ua.edu/fermi/}, accessed on 06/06/2023).}
\label{fig:1}
\end{figure*}

The automated pipeline, illustrated in Fig. \ref{fig:2}, listens for incoming AMON\_ICECUBE\_GOLD and AMON\_ICECUBE\_BRONZE event alerts coming in through the GCN. These are individual single-neutrino events in the IceCube detector with a neutrino energy in the sub-PeV to 1 PeV energy regime and are shown in Fig.~\ref{fig:1} for the time interval between June 19, 2019 and June 6, 2023. 

More specifically, the analysis pipeline extracts information from the GCN alert related to the type of event (GOLD or BRONZE),  the sky coordinates of the alert along with the associated confidence interval, the date and time of the alert in UTC, and also the corresponding \emph{Run\_ID} and \emph{Event\_ID}. The main components of the tool include performing an automatic \textit{Fermi}-LAT analysis of the neutrino RoI, calculating the visibilities for common follow-up instruments and collecting multiwavelength archival data for known sources in the RoI. The entire process is then repeated for multiple revisions of a particular alert.

The tool then runs an automatic analysis of the \textit{Fermi}-LAT photons detected from within the RoI of the neutrino alert over a 30 day interval prior to each individual event. Each analysis uses the \textit{Fermi} Science Tools version $11-05-03$~\footnote{\url{http://fermi.gsfc.nasa.gov/ssc/data/analysis/software} (accessed on 06/06/2023)}, \textit{FERMIPY} version 1.0.1~\footnote{\url{http://fermipy.readthedocs.io} (accessed on 06/06/2023)} \citep{wood2017fermipy} in conjunction with the \textit{PASS} 8 instrument response functions~\citep{atwood2013pass}. The contributions from the isotropic and Galactic diffuse backgrounds are modeled using the most recent templates for isotropic and Galactic diffuse emission, iso\_P8R3\_SOURCE\_V2\_v1.txt and gll\_iem\_v07.fits respectively. Sources in the 4FGL-DR3 catalog \citep{4fgl_dr3} within a radius of $20^{\circ}$ from the best-fit location of each alert are included in the model with their spectral parameters fixed to their catalog values. The \textit{gtfindsrc} routine is also applied to search for any additional point sources not accounted for in the model. Any source found to have a test statistic (TS, \cite{RN7}) $\geq$ 9 (roughly corresponding to a significance of $\sim$ 3$\sigma$) is permanently added to the model at the position of its highest TS value.


Moreover, the normalization factor for both the isotropic and Galactic diffuse emission templates are left free to vary, along with the spectral normalization of all modeled sources within the RoI. Finally, the spectral shape parameters of all modeled sources within 3$^{\circ}$ of the alert are left free to vary while those of the remaining sources are fixed to the values reported in the 4FGL-DR3 catalog. 
A binned likelihood analysis is then performed in order to obtain the spectral parameters best describing the model during the period of observation, using a spatial binning of 0.1$^{\circ}$ pixel$^{-1}$ and two energy bins per decade. 
The results, including HTML, FITS files, and images of plots, are mirrored on to the web server (\url{https://multimessenger.ua.edu/fermi/}, accessed on 06/06/2023).


\section{Discussion}
\label{sec:discussion}

\begin{figure*}[t!]
\centering
\includegraphics[width=  0.48 \linewidth]{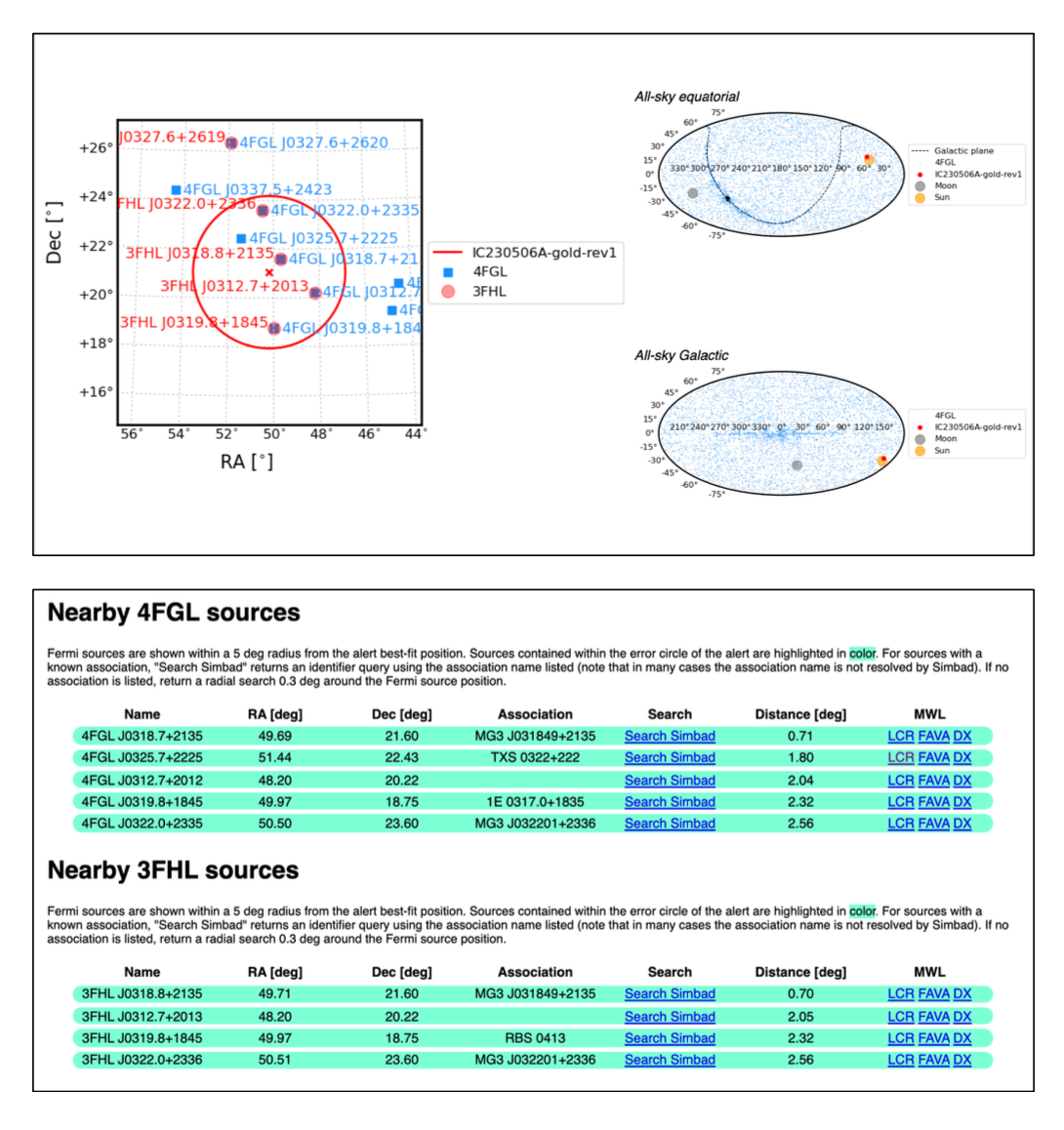}
\includegraphics[width=  0.39 \linewidth]{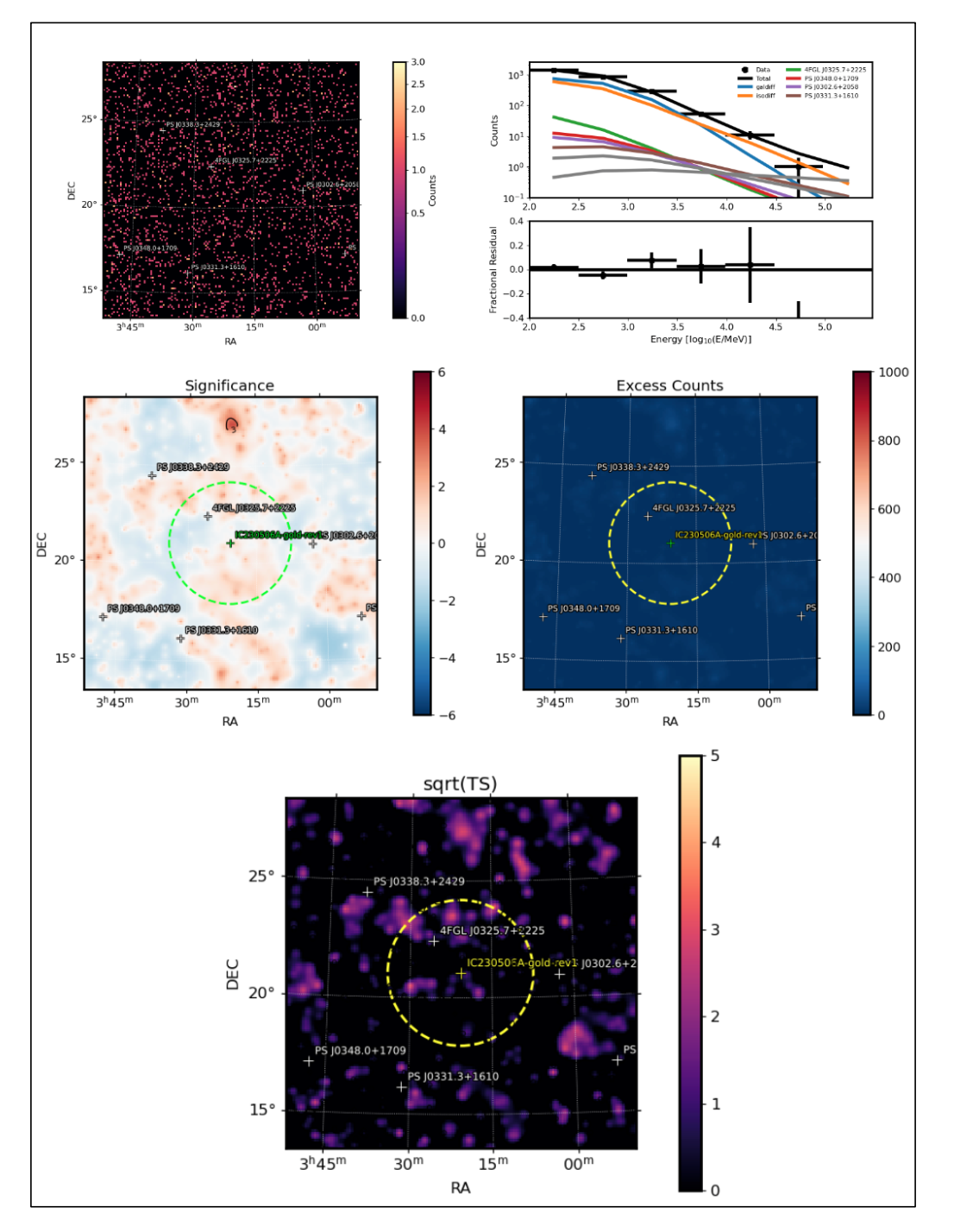}
\caption{\textbf{Left: Top panel:} A skymap of the neutrino alert, also containing all sources in the confidence region from the 4FGL-DR3 and 3FHL catalog. \textbf{Bottom panel:} These sources, listed separately, with links to the source entry in \emph{Simbad}, as well as multiwavelength archival data repositories. \textbf{Right:} The diagnostic plots produced in this \emph{Fermi}-LAT data analysis. These include a skymap of the residual significance and the excess photon counts obtained in the analysis over the RoI centred on the location of the alert. The colour scales correspond to the excess significance of each pixel in the RoI in Gaussian $\sigma$ and the number of excess photons at each pixel in the RoI respectively.}
\label{fig:3}
\end{figure*}

\begin{figure*}[t!]
\centering
\includegraphics[width=  0.47 \linewidth]{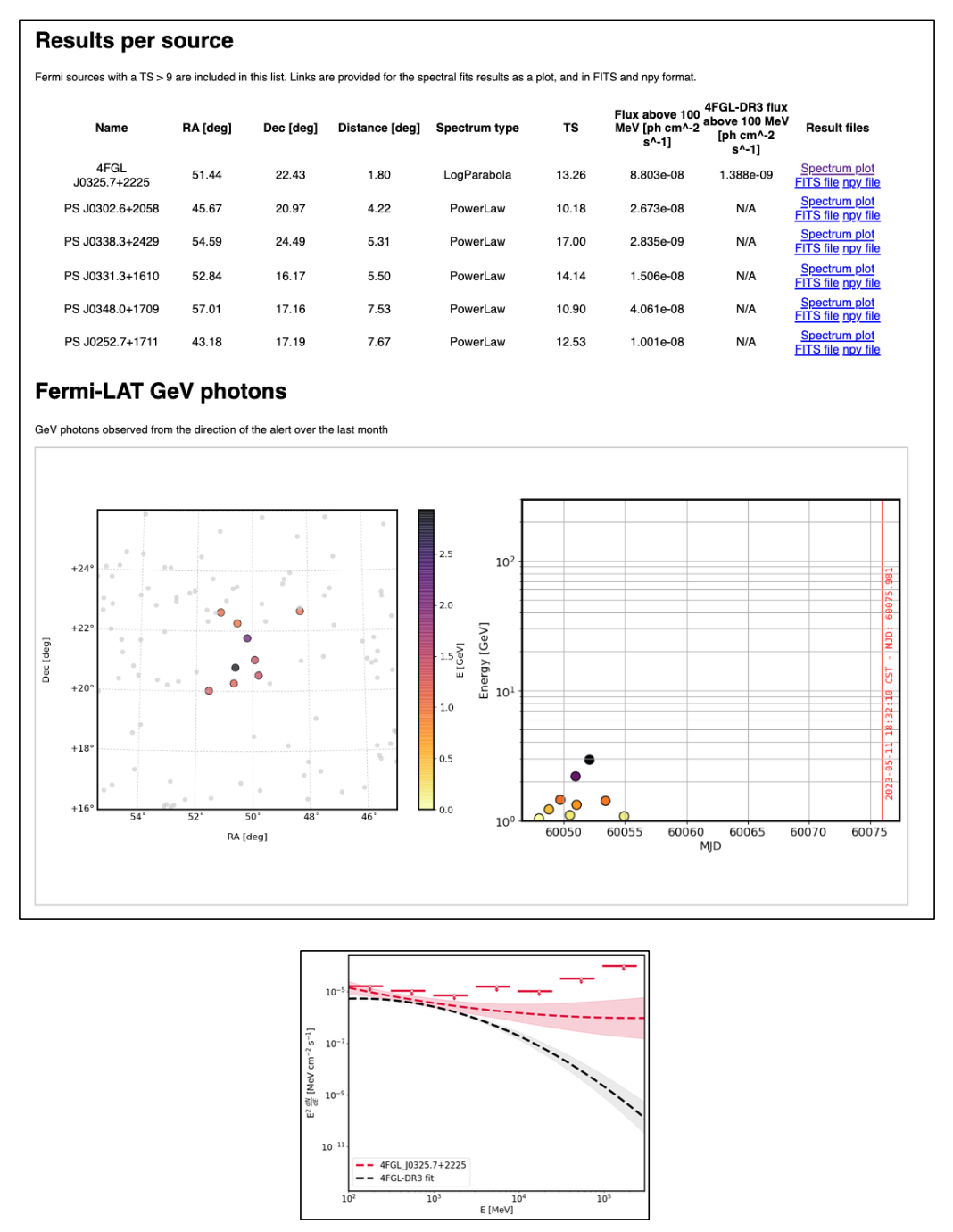}
\includegraphics[width=  0.37 \linewidth]{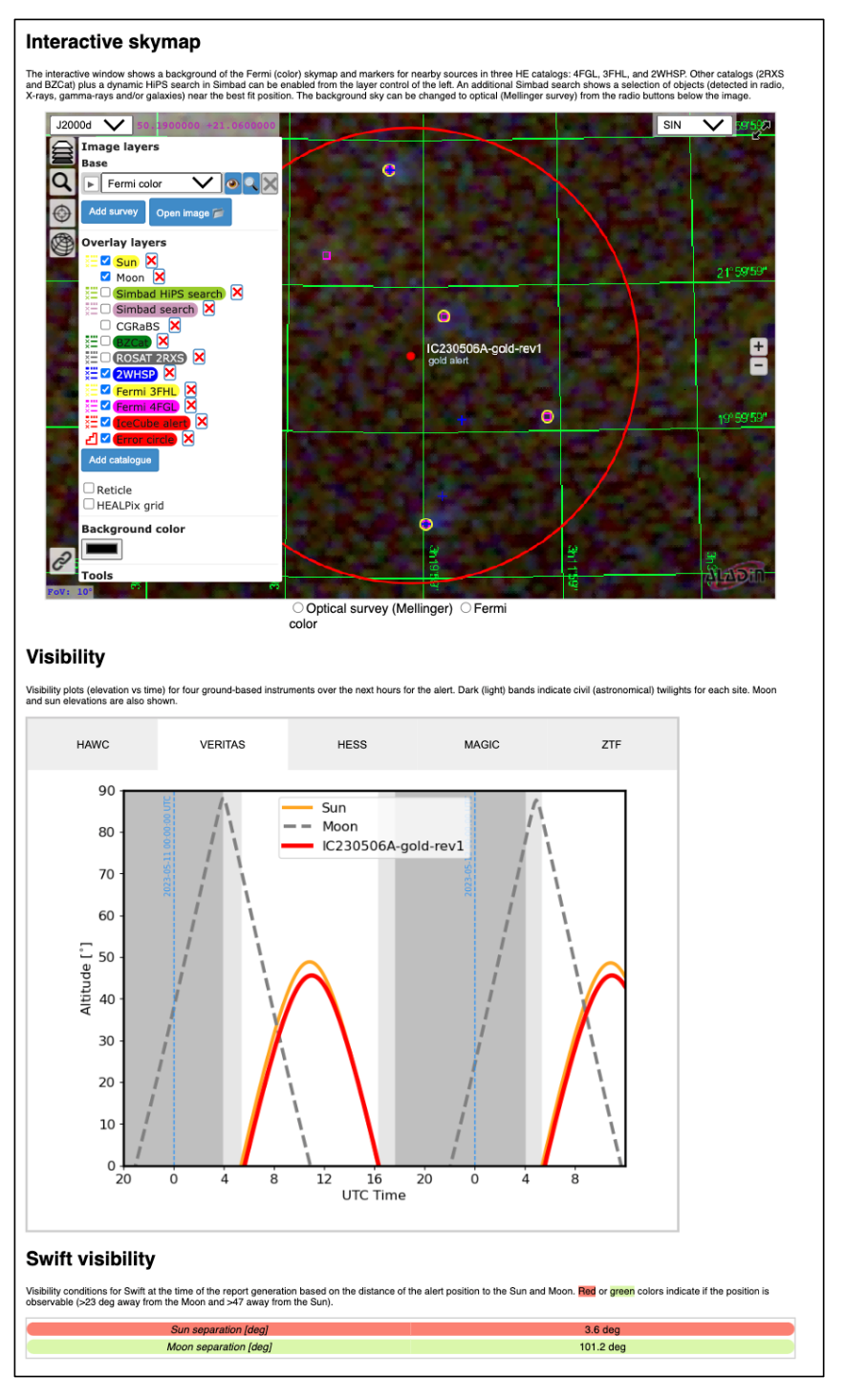}
\caption{\textbf{Left: Top panel:} The results obtained from the \emph{Fermi}-LAT analysis, including the TS, flux and spectrum for both known 4FGL sources, as well as potential new sources. \textbf{Middle panel:} A skymap and lightcurve of the GeV photons emitted from within the RoI. \textbf{Bottom panel:} An example of a \emph{Fermi}-LAT spectrum for the source 4FGL J0325.7+2225 in red. The data are binned into two energy bins per decade, with individual bins having a TS $<$ 4 (roughly corresponding to a significance of $\sim$ 2$\sigma$) considered as upper limits. Also shown for comparison is the 4FGL-DR3 best-fit spectrum. \textbf{Right: Top panel:} An interactive skymap containing markers for nearby sources in the 4FGL, 3FHL, and 2WHSP catalog. \textbf{Bottom panel:} A visibility plot of the RoI with VERITAS.}
\label{fig:4}
\end{figure*}

In this Section, we consider a typical example of a neutrino alert, IC230506A-gold-rev1\footnote{\url{https://gcn.gsfc.nasa.gov/notices_amon_g_b/137910_29871391.amon} (accessed on 06/06/2023)}, received on 2023-05-06 at 15:53:45 UTC. This was a revised AMON\_ICECUBE\_GOLD type alert, with \emph{Run\_ID} = 137910, \emph{Event\_ID} = 29871391, and the associated sky coordinates are RA~=~$50.19^{\circ}$, Dec = $21.06^{\circ}$, with a corresponding error radius of $3.12^{\circ}$. The plots produced by the automated tool are shown in  Fig. \ref{fig:3} and Fig. \ref{fig:4}.

The output results include a skymap of the neutrino alert, also containing all sources in the confidence region from the 12-year \emph{Fermi}-LAT catalog (4FGL-DR3, \cite{4fgl_dr3}), as well as the third \emph{Fermi}-LAT catalog  of sources significantly detected in the 10 GeV -- 2 TeV energy range (3FHL, \cite{3fhl}). These sources are also listed separately in a Table containing their 4FGL and 3FHL names, sky coordinates, distance from the neutrino alert and links to the source entry in \emph{Simbad} \citep{2000_simbad}, as well as multiwavelength archival data repositories including the \emph{Fermi}-LAT Light Curve Repository \citep{2023_lcr} and the \emph{Fermi} All-sky Variability Analysis (FAVA, \cite{2017_fava}) Light Curve Repository.

Also shown in Fig. \ref{fig:3}, are the set of diagnostic plots produced in this \emph{Fermi}-LAT data analysis. These are obtained for each individual alert and subsequent revision and include a skymap showing the modelled distribution of the gamma-ray photons obtained after removing all sources found to have a TS~<~9 followed by applying the \textit{gtfindsrc} routine over the RoI, as well as a skymap of the residual significance and the excess photon counts obtained in the analysis over the RoI centred on the location of the alert.

Fig. \ref{fig:4} shows more results obtained from the \emph{Fermi}-LAT analysis, including both known 4FGL sources having TS $\geq 9$, as well as new sources obtained from the \textit{gtfindsrc} routine. This includes the TS obtained for each source, the measured flux values in the energy range 100 MeV -- 300 GeV along with the corresponding 4FGL flux values for known sources. There is also a link to corresponding plots, such as the one shown in Fig. \ref{fig:4}, of the \emph{Fermi}-LAT spectrum for each source, with the 4FGL-DR3 best-fit spectrum  \citep{4fgl_dr3} also shown for comparison for known sources. Furthermore, we also obtain a skymap and lightcurve of the GeV photons, emitted from within the RoI and having a $\geq$~99~$\%$ probability of originating from each individual source, over the one month observation period investigated. 

Moreover, the tool also produces an interactive skymap containing markers for nearby sources in three catalogs, namely the 4FGL, 3FHL, and 2WHSP \citep{2017_chang}. Other catalogs, for example the 2RXS \citep{2016_rxs} and BZCat \citep{2009_BZ} plus a dynamic HiPS search in \emph{Simbad}, can also be enabled. An additional \emph{Simbad} search can be made to show a selection of objects detected in radio, X-rays, gamma-rays and galaxies near the best fit position of the alert. The background sky can also be changed to optical (Mellinger survey, \citep{2009_Mellinger}) using the radio buttons below the image. Finally, we also produce visibility plots of the RoI for the four ground-based instruments, namely MAGIC, H.E.S.S., VERITAS and HAWC, over the time interval just after the alert, in order to help enable prompt follow-up observations.

\section{Conclusions and Future Work}
\label{sec:conclusions}

In this work, we introduce an automated tool that aims at using \emph{Fermi}-LAT data to identify multiwavelength counterparts to astrophysical neutrino events and enable prompt follow-up observations with ground-based and space-based observatories in order to help pinpoint the neutrino source. After discussing the main components of the the analysis and processing pipeline, we walk-through the primary outputs from the analysis tool for a typical example of a neutrino alert, IC230506A-gold-rev1.

It should be noted that this is an early version of the automated tool with plenty of scope for even further improvement. This includes, for example, adding support for 1-year and full-mission \emph{Fermi}-LAT analysis alongside the month long time period currently analyzed, gathering more multiwavelength data and resources to enable SED construction, and finally improving compatibility with the GCN Kafka Client setup \footnote{\url{https://gcn.nasa.gov/docs/client} (accessed on 06/06/2023)}.

\begin{acknowledgments}

A.A. and M.S. acknowledge support through NASA grants 80NSSC22K1515, 80NSSC22K0950, 80NSSC20K1587, 80NSSC20K1494 and NSF grant PHY-1914579. 

\end{acknowledgments}

\bibliography{references}{}
\bibliographystyle{JHEP}


%
%
%

\end{document}